\documentstyle[preprint,aps]{revtex}
\begin{document}
\draft
\title{Dimensional renormalization in $\phi ^3$ theory: ladders and rainbows}
\author{R.~Delbourgo\cite{Author1} and D.~Elliott\cite{Author2}}
\address{University of Tasmania, GPO Box 252-21, Hobart,\\
Tasmania 7001, Australia}
\author{D.S.~McAnally\cite{Author3}}
\address{University of Queensland, St Lucia, Brisbane, \\
Queensland 4067, Australia}
\date{\today }
\maketitle

\begin{abstract}
The sum of all the ladder and rainbow diagrams in $\phi ^3$ theory near 6
dimensions leads to self-consistent higher order differential equations in
coordinate space which are not particularly simple for arbitrary dimension $%
D $. We have now succeeded in solving these equations, expressing the
results in terms of generalized hypergeometric functions; the expansion and
representation of these functions can then be used to prove the absence of
renormalization factors which are transcendental for this theory and this
topology to all orders in perturbation theory. The correct anomalous scaling
dimensions of the Green functions are also obtained in the six-dimensional
limit.
\end{abstract}


\narrowtext

\section{INTRODUCTION}


In a recent paper\cite{RDAKGT} we managed to derive closed forms for ladder
corrections to self-energy graphs (rainbows) and vertices, in the context of
dimensional renormalization. We only succeeded in carrying out this program
for Yukawa couplings near four dimensions, although we did obtain the
differential equations pertaining to the $\phi^3$ theory as well; but we
were not able to solve the latter in simple terms. We have now managed to
obtain closed expressions for $\phi^3$ theory as well and wish to report the
results here. The answers are indeed non-trivial and take the form of $_0F_3$
functions, which perhaps explains why they had eluded us so far.
Interestingly, the closed form results for Yukawa-type models lead to Bessel
functions with curious indices and arguments; but as these can also be
written as $_0F_1$ functions, the analogy with $\phi^3$ is close after all.

Given the exact form of the results, both for rainbows and ladders, we are
able to test out Kreimer's\cite{DK} hypothesis about the connection between
knot theory and renormalization theory with confidence, fully verifying that
the renormalization factors for such topologies are indeed
non-transcendental. At the same time we are able to determine the $Z$%
-factors to any given order in perturbation theory and show that in the $D
\rightarrow 6$ limit, the correct anomalous dimensions of the Green
functions do emerge, which is rather satisfying. The various $Z$-factors
come out as poles in $1/(D-6)$ when the Green functions are expanded in the
normal way as powers of the coupling constant, but the complete result
produces the renormalized Green function to all orders in coupling for any
dimension $D$.

In the next section we treat the vertex diagrams, converting the
differential equation for ladders into hypergeometric form. Upon picking the
correct solution we are able to do two things: (i) establish that in the $D
\rightarrow 6$ limit one arrives at the correct anomalous scaling factor for
the vertex function, and (ii) obtain the $Z$-factors through a perturbative
expansion of argument of the hypergeometric function. The case (i) is a bit
tricky; it requires an asymptotic analysis, because the indices of the
hypergeometric function as well as the argument diverge in the
six-dimensional limit. The next section contains the analysis of the rainbow
graphs; the equations are similar to the vertex case, but different
solutions must be selected, resulting in a different anomalous dimension. It
is nevertheless true that the self-energy renormalization constant remains
non-transcendental. A brief concluding section ends the paper.

\section{Ladder vertex diagrams}


We will only treat the massless case, since this is sufficient to specify
the $Z$-factors once an external momentum scale is introduced. To further
simplify the problem we shall consider the case where the vertex is at
zero-momentum transfer, leaving just one external momentum $p$. The equation
for the 1-particle irreducible vertex $\Gamma $, in the ladder
approximation, thereby reduces to 
\begin{equation}
\Gamma (p)=Z+ig^2\int \frac 1{q^2}\Gamma (q)\frac 1{q^2}\frac{d^Dq/(2\pi )^D%
}{(p-q)^2}.
\end{equation}
Letting $\Gamma (p)\equiv p^4G(p)$, the equation can be Fourier-transformed
into the coordinate-space equation for G, 
\begin{equation}
\lbrack \partial ^4-ig^2\Delta _c(x)]G(x)=Z\delta ^D(x),
\end{equation}
where $\Delta _c$ is the causal Feynman propagator for arbitrary dimension $%
D $. Since the coupling $g$ is dimensionful when $D\neq 6$, it is convenient
to introduce a mass scale $\mu $ and define a dimensionless coupling
parameter $a$ via, 
\[
\frac{g^2}{4\pi ^{D/2}}\frac{\Gamma (D/2-1)}{(-x^2)^{1-D/2}}\equiv \frac{%
4a(\mu r)^{6-D}}{r^4}. 
\]
Then, rotating to Euclidean space $(r^2=-x^2)$, the ladder vertex equation
simplifies to 
\begin{equation}
\left[ (\frac{d^2}{dr^2}+\frac{D-1}r\frac d{dr})^2-\frac{4a(\mu r)^{6-D}}{r^4%
}\right] G(r)=Z\delta ^D(r).
\end{equation}
This is trivial to solve when $D=6$, since it becomes homogeneous for $r>0$
and the appropriate solution is 
\[
G(r)\propto r^b;\qquad b=-1-\sqrt{5-2\sqrt{4+a}}, 
\]
reducing to $G(r)\propto r^{-2}$ or $\Gamma (p)=1$ in the free field case ($%
a=0$); it represents a useful limit when analysing the full equation (3), to
which we now turn.

Let us define the scaling operator $\Theta _r=r\frac d{dr}$. This allows us
to rewrite the square of the d'Alembertian operator as 
\begin{equation}
\partial ^4=\left[ \frac{d^2}{dr^2}+\frac{D-1}r\frac d{dr}\right]
^2=r^{-4}(\Theta _r-2)\Theta _r(\Theta _r+D-4)(\Theta _r+D-2).
\end{equation}
Hence for $r>0$ the original equation (3) reduces to the simpler form, 
\begin{equation}
\left[ \Theta _\rho (\Theta _\rho -2)(\Theta _\rho +D-4)(\Theta _\rho
+D-2)-4a\rho ^{6-D}\right] G=0,
\end{equation}
where $\rho =\mu r$ and $\Theta _\rho $ is the corresponding scaling
operator. Next, rescaling the argument to $t=4a\nu ^4\rho ^{-1/\nu },$ with $%
\nu \equiv 1/(D-6)$, we obtain the hypergeometric equation: 
\begin{equation}
\left[ \Theta _t(\Theta _t+2\nu )(\Theta _t-1-2\nu )(\Theta _t-1-4\nu
)-t\right] G=0.
\end{equation}
Being of fourth order, there are four linearly independent solutions 
\[
_0F_3(b_1,b_2,b_3;t), 
\]
\[
{t^{1-b_1}}_0F_3(2-b_1,b_2-b_1+1,b_3-b_1+1;t) 
\]
\[
{t^{1-b_2}}_0F_3(2-b_2,b_3-b_2+1,b_1-b_2+1;t), 
\]
\[
{t^{1-b_3}}_0F_3(2-b_3,b_1-b_3+1,b_2-b_3+1;t), 
\]
where $b_1\equiv 1+2\nu ,\quad b_2=-4\nu ,\quad b_3=-2\nu $. The appropriate
solution, which near $t=0$ behaves as $r^{4-D}$ when $a=0$ , is the last
choice, namely 
\[
G\propto {t^{1+2\nu }}_0F_3(2+2\nu ,2+4\nu ,1-2\nu ;t). 
\]
Near $r=0$ this behaves like $r^{-2-1/\nu }$. Finally, renormalizing the
Green function $G$ to equal $\mu ^{D-4}$ when $r=1/\mu $, the scale we
introduced previously for the coupling constant, and restoring the original
variables, we end up with the exact result 
\begin{equation}
G(r)=r^{4-D}\frac{_0F_3(2-\frac 2{6-D},2-\frac 4{6-D},1+\frac 2{6-D};\frac{%
4a(\mu r)^{6-D}}{(6-D)^4})}{_0F_3(2-\frac 2{6-D},2-\frac 4{6-D},1+\frac 2{6-D%
};\frac{4a}{(6-D)^4})}.
\end{equation}
To check that the poles in $(D-6)$ cancel out at any given order in
perturbation theory, one simply expands the numerator and denominator in (7)
to any particular power in the dimensionless coupling $a$ and take the limit
as $D\rightarrow 6$. For instance, to order $a^3$, with a little work one
arrives at, 
\begin{equation}
r^2G(r)\rightarrow 1+\frac a4\ln (\mu r)+\frac{a^2}{64}\ln (\mu r)\left(
1+2\ln (\mu r)\right) +\frac{a^3}{1536}\ln (\mu r)\left( 9+6\ln (\mu r)+4\ln
^2(\mu r)\right) +O(a^4).
\end{equation}
It is most gratifying that this agrees perfectly with the expansion of the
scaling index $b$ obtained previously at $D=6$. The most significant point
is that there is no sign of a transcendental constant in the singularities
of the perturbation expansion for $\phi ^3$ theory near 6-dimensions,
signifying that the renormalization constant $Z$ is free of them, in
agreement with the Kreimer hypothesis based on knot theory.

One last (rather difficult) check on our work is to see what happens
directly to (7) as $D$ approaches 6, without having to invoke perturbation
theory. For that an asymptotic analysis\cite{f1} based on the method of
steepest descent (see for example, de Bruijn\cite{DB}) is needed.
We start by making use of the Barnes integral representation of
the hypergeometric function, 
\[
_0F_3(b_1,b_2,b_3;t)=\frac 1{2i\pi }\int_{-i\infty }^{+i\infty }\frac{\Gamma
(b_1)\Gamma (b_2)\Gamma (b_3)}{\Gamma (b_1+z)\Gamma (b_2+z)\Gamma (b_3+z)}%
\Gamma (-z)t^z\,dz.
\]
In our case the $b$ arguments lead us to evaluate the integral 
\begin{equation}
I_\nu (r)\equiv \frac 1{2i\pi }\int_{-i\infty }^{i\infty }\frac{\Gamma
(2+4\nu )\Gamma (2+2\nu )\Gamma (1-2\nu )\Gamma (-z)}{\Gamma (2+4\nu
+z)\Gamma (2+2\nu +z)\Gamma (1-2\nu +z)}[4a\nu ^4\rho ^{-1/\nu }]^z\,dz
\end{equation}
in the limit as $\nu \rightarrow \infty $. We shall show that as a function
of $\rho $,  $I_\nu$ behaves like $\rho ^{1-\sqrt{5-2\sqrt{4+a}}%
}$. Remember that $G(r)\propto r^{-2-1/\nu }I_{\nu}(r)$.

For the method of steepest descents, suppose we write $I_\nu $ as 
\[
\frac 1{2\pi i}\int_{-i\infty}^{i\infty} g_\nu (z)\exp [{f}_\nu {(z)]}\,dz
\]
where $g_\nu $ is a ``slowly varying'' function. We see from (9) that all
the poles of the integrand lie on the positive real axis. If $\zeta $ is
such that Re($\zeta $)$<0$ and $f_\nu ^{\prime }(\zeta )=0$, then an
approximate evaluation of $I_\nu $ is given by
\[
\alpha g_{\nu}(\zeta)\exp[f_{\nu}(\zeta)]/\sqrt{2\pi
 |f_{\nu}^{\prime \prime }(\zeta )|},
\]
where
\[
\alpha \equiv \exp [-i\arg (f_\nu ^{\prime \prime }(\zeta ))/2].
\]
On applying the reflection formula for the gamma function\cite{AS} to both
$\Gamma(1-2\nu )$ and $\Gamma (1-2\nu +z)$ appearing in (9), we find that
we can write 
\[
g_\nu (z)=\rho ^{-z/\nu }\frac{\sin \pi (2\nu -z)}{\sin (2\pi \nu )},
\]
provided $\nu $ is not an integer, and 
\[
\exp [f_\nu (z)]=\frac{\Gamma (2+4\nu )\Gamma (2+2\nu )\Gamma (2\nu
-z)\Gamma (-z)}{\Gamma (2+4\nu +z)\Gamma (2+2\nu +z)\Gamma (2\nu )}(4a\nu
^4)^z.
\]
Since 
\[
f_\nu ^{\prime }(z)=\log (4a\nu ^4)-[\psi (2\nu -z)+\psi (-z)+\psi (2+2\nu
+z)+\psi (2+4\nu +z)],
\]
where $\psi $ denotes the psi (or digamma) function, we look for a zero at $%
z=-\xi \nu $ say, where  $0<$ $\xi <2$. Since we assume $\nu \gg 1$ and
since for $x\gg 1$, $\psi (x)=\log x+O(1/x),$ we find that $\xi $ must
satisfy the quartic  
\[
\xi (\xi +2)(\xi -2)(\xi -4)=4a .
\]
The four solutions of this equation are
\[
\xi =1\pm \sqrt{5\pm 2\sqrt{4+a}}
\]
and are all real if $0\leq a<9/4.$ In particular we shall choose the zero $%
\beta $ say in (0,2) which is closest to the origin; that is
\[
\beta =1-\sqrt{5-2\sqrt{4+a}}.
\]
With this value of $\beta $ we find
\[
f_\nu^{\prime\prime}(-\beta\nu )\simeq (1-\beta)(4+2\beta -\beta ^2)/(a\nu ).
\]
Since in fact $0<\beta <1,$ we have that $\arg f_\nu^{\prime\prime }(-\beta\nu
)=0$ so that $\alpha =1$. Again,
\[
g_\nu (-\beta \nu )=\frac{\sin ((2+\beta )\pi \nu )}{\sin (2\pi \nu )}\rho
^\beta 
\]
and, after some algebra,
\[
\exp [f_\nu (-\beta \nu )]=\frac{4\sqrt{2\pi }\beta (\beta +2)}{a^{3/2}\nu
^{1/2}}[\frac{16(2+\beta )}{(4-\beta )^2(2-\beta )}]^{2\nu }\exp (-4\beta
\nu ),
\]
approximately. Consequently, for $\nu \gg 1$ but not an integer, we find
\[
_0F_3(2+2\nu ,2+4\nu ,1-2\nu ;4a\nu ^4\rho ^{-1/\nu })\sim \frac{\rho ^\beta
\sin ((2+\beta )\pi \nu )}{a\sin (2\pi \nu )}\frac{4\beta (\beta +2)\exp(
-4\beta\nu)}{
(1-\beta )^{1/2}(4+2\beta -\beta ^2)^{1/2}}[\frac{16(2+\beta )}{(4-\beta
)^2(2-\beta )}]^{2\nu }
\]

Using this asymptotic expansion, we obtain simply from eq.(7) that  
\begin{equation}
G(r)=\mu ^\beta r^{5-D-\sqrt{5-2\sqrt{4+a}}},
\end{equation}
which is just the scaling behaviour at 6-dimensions which we were seeking.
We have therefore fully verified the correctness of (7) in all the limits.
The last step is to convert the answer to Minkowski space by making the
familiar substitution $r^2\rightarrow -x^2+i\epsilon .$

\section{Rainbow Diagrams}

Let $\Delta_R(p)$ denote the renormalized $\phi$ propagator in rainbow
approximation, so that $p^2\Delta_R(p)=1-\Sigma_R(p)/p^2$, where $\Sigma_R$
is the rainbow self-energy. The propagator obeys the integral equation in
momentum space 
\begin{equation}
p^4\Delta_R(p) = Zp^2 + ig^2\int \frac{d^Dk}{(2\pi)^D}\frac{\Delta_R(p-k)}{%
k^2},
\end{equation}
where $Z$ now refers to the wave-function renormalization constant. As
always we convert this into an $x$-space differential equation, 
\begin{equation}
[\partial^4 -ig^2\Delta_c(x)]\Delta_R(x) = -Z\partial^2\delta^D(x).
\end{equation}
Interestingly, this is exactly the same equation as (2), apart from the
right hand side, and it can therefore be converted into hypergeometric form
by following the same steps as before. The only difference is that we should
look for a different solution, because as $g\rightarrow 0$, $%
\Delta_R(p)\rightarrow 1/p^2$, or $\Delta_R(x) \sim (x^2)^{1-D/2}.$

A simple analysis shows the correct solution is 
\[
{t^{1+4\nu}}_0F_3(2+4\nu,2+6\nu,1+2\nu;t);\qquad t=4a\nu^4(\mu r)^{-1/\nu},
\quad \nu=1/(D-6), 
\]
because this reduces to $r^{2-D}$ when $a=0$. Actually we can solve (12)
directly at $D=6$ when $a \neq 0$ because it is a simple homogeneous
equation leading to 
\[
\Delta_R(x) \propto r^{-1-\sqrt{5+2\sqrt{4+a}}} 
\]
and thereby determine the anomalous dimension from the exponent of $r$.
Anyhow, the exact solution of the rainbow sum for any $D$ and renormalized
at $r=1/\mu$ is here obtained to be 
\begin{equation}
\Delta_R(r) = r^{2-D}\frac{_0F_3(2-\frac{4}{6-D},2-\frac{6}{6-D}, 1-\frac{2}{%
6-D};\frac{4a(\mu r)^{6-D}}{(6-D)^4})} {_0F_3(2-\frac{4}{6-D},2-\frac{6}{6-D}%
,1-\frac{2}{6-D}; \frac{4a}{(6-D)^4})}.
\end{equation}
The numerator and denominator of (13), when expanded in powers of $a$ will
reproduce the (renormalized) perturbation series; to third order we find, in
the limit as $D \rightarrow 6$, that all poles disappear and 
\[
r^4\Delta_R(r) = 1 -\frac{a}{12}\ln(\mu r) + \frac{a^2}{1728}\ln(\mu r) (11
+ 6\ln(\mu r)) - \frac{a^3}{124416}\ln(\mu r) (103 + 66\ln(\mu r) +
12\ln^2(\mu r)) + O(a^4). 
\]
This coincides perfectly with the expansion of the scaling exponent at $D=6$.

Lastly we need to show that the $D \rightarrow 6$ limit of (13) collapses to
the scaling behaviour found above, via an asymptotic analysis of the Barnes
representation. We have indicated how this can be proven in the previous
section and thus we skip the formal details to avoid boring the reader. The
long and the short of the analysis is that no transcendentals enter into the
above expressions for the self-energy (including their singularities, which
are tied to the wave-function renormalization constant). These results
confirm nicely the Kreimer\cite{DK} hypothesis that the $Z$-factors will be
simple rationals for such topologies.

\section{Conclusions}


We have succeeded in evaluating an all-orders solution of Green functions for
ladder and rainbow diagrams for any dimension $D$ in $\phi^3$ theory; the
results are non-trivial, involving $_0F_3$ hypergeometric functions. We have
demonstrated that, in the limit as $D \rightarrow 6$, the correct
six-dimensional scaling behaviour (which can be separately worked out) is
reproduced. One can likewise determine the exact solutions for massless
bubble ladder exchange in $\phi^4$ theory, because the equations are very
similar: they are also of fourth order and can be converted into
hypergeometric form too\cite{DBDK}.

More intriguing is the question of what happens when self-energy and ladder
insertions are considered, so far as renormalization constants are
concerned. A recent paper by Kreimer\cite{DK2} has shown that such
topologies with their disjoint divergences can produce transcendental $Z$ in
accordance with link diagrams that are of the (2,$q$) torus knot variety,
where the highest $q$ is determined by the loop number. It would be
interesting to show this result without resorting to perturbation theory by
summing all those graphs exactly, as we have done in this paper. (Kreimer
cautions that multiplicative renormalization may screen his new findings.)
The generalization to massive propagators\cite{SE} does not seem beyond the
realms of possibility either, although it has a marginal bearing on $Z$%
-factors.

\acknowledgments
We thank the University of Tasmania Research Committee for providing a small 
grant during 1996 which enabled this collaboration to take place.


\begin{references}
\bibitem{f1}  In reference 1, we deduced the Debye expansion of the Bessel
function (and a variant of it) by this artifice, adopting a particularly
helpful integral representation of Bessel function. In this paper we are
obliged to use the Barnes representation as it is the only version which
readily generalizes to $_pF_q$ hypergeometric functions; the stationary
point analysis is completely different now and has been included for that
reason.
\end{references}
\end{document}